\documentclass[twocolumn,preprintnumbers,showpacs,amsmath,amssymb]{revtex4}
\usepackage{graphicx}
\usepackage{amssymb}

\begin{document}
\title{Security limitation on a class of device-independent quantum key
distribution}
\author{Guang Ping He}
\email{hegp@mail.sysu.edu.cn}
\affiliation{School of Physics and Engineering, Sun Yat-sen University, Guangzhou 510275,
China}

\begin{abstract}
Recently there were many proposals on device-independent (DI) quantum key
distribution protocol whose security is based on the violation of the
Clauser-Horne-Shimony-Holt inequality. However, as a statistical law, a
certain extent of fluctuation has to be allowed. We show that the
eavesdropper can make use of this property to obtain a remarkable part of
the secret key by replacing some of the DI nonlocal boxes
with local ones. On the contrary, the same cheating strategy does not apply
to the device-dependent (DD) version of the protocol. Thus such kind of DI protocol
is less secure than its DD counterpart.
\end{abstract}

\maketitle

\newpage


\section{Introduction}

Device-independent (DI) quantum cryptography has caught great interests
recently \cite{qbc29,di22,qi994,di15,di6,di8,di18,di20,di11,di25}. It aims
to replace the model of the physical devices used in the cryptographic
protocol with physically testable assumptions, e.g., the certification of
nonlocality. Thus the devices can be treated as black boxes that produce
outputs correlated with some inputs. This brings the advantage that the
assumptions needed to guarantee the security of the protocol can be
significantly reduced, so that the knowledge of the internal workings of the
devices is not required. The protocol remains reliable even if the devices
are provided by the adversary. Such a higher degree of security makes DI
protocols more dependable in practical applications than traditional quantum
cryptography.

In many recent proposals on DI quantum key distribution (QKD) protocol \cite%
{qi1065,qi481,qi996,qi997,qi998,di3,qi992}, the nonlocality is tested by
observing the violation of the Clauser-Horne-Shimony-Holt (CHSH) inequality.
However, being a statistic law, fluctuation deviated from the expected CHSH
value is inevitable in principle, even if experimental imperfections are not
taken into account. Here we show that it leaves rooms for the
eavesdropper to use local boxes in place of the DI boxes with a certain frequency, so
that he stands a non-trivial probability to learn a part of the raw
key while escaping the detection. On the other hand, we will also show that the device-dependent (DD) version of the same protocol is completely immune to this
specific cheating strategy. This result indicates that such a DI QKD
protocol is less secure than its own device-dependent (DD) version.

We will see that this security limitation is a fundamental theoretical problem
of DI QKD, which is not caused by the imperfections of experimental devices.
Therefore for simplicity, such imperfections will not be considered here.
That is, we only study the ideal case without transmission errors, detection
loss, or dark counts, etc.

In order to make it easier to understand the DI QKD protocol based on the
CHSH inequality, we will first describe its DD version in the next section.
Then we give the DI one in section III. In section IV we will explain which
kind of statistical fluctuation is considered in this paper. The cheating
strategy on DI QKD will be elaborated in section V, and it will be shown in
section VI why the strategy does not work in the DD protocol. Finally, we
summarize our result in section VII.

\section{The DD QKD protocol}

Following the notations of \cite{qi481,qi996}, a particular DD
implementation of the QKD protocols based on the CHSH inequality can be accomplished
in the following manner. Alice and Bob share a quantum channel consisting
of a source that supplies many pairs of entangled particles, each of which is
supposed to be in the Bell state%
\begin{equation}
\left\vert \Phi ^{+}\right\rangle =(\left\vert 0\right\rangle _{A}\left\vert
0\right\rangle _{B}+\left\vert 1\right\rangle _{A}\left\vert 1\right\rangle
_{B})/\sqrt{2},  \label{Bell}
\end{equation}%
where particle $A$ ($B$) goes to Alice (Bob), with $\left\vert 0\right\rangle $
and $\left\vert 1\right\rangle $ denoting the two eigenstates of the Pauli
operator $\sigma _{z}$. On each of their particles, Alice chooses randomly
among three measurements%
\begin{eqnarray}
A_{0} &=&\sigma _{z},  \nonumber \\
A_{1} &=&(\sigma _{z}+\sigma _{x})/\sqrt{2},  \nonumber \\
A_{2} &=&(\sigma _{z}-\sigma _{x})/\sqrt{2},  \label{Alice}
\end{eqnarray}%
while Bob chooses randomly between two measurements%
\begin{eqnarray}
B_{1} &=&\sigma _{z},  \nonumber \\
B_{2} &=&\sigma _{x}.  \label{Bob}
\end{eqnarray}%
The outcomes of the measurements are labeled as $a_{i},b_{j}\in \{+1,-1\}$ ($%
i=0,1,2$, $j=1,2$). Alice and Bob announce their inputs (i.e., their choices of
the measurements) through a classical
channel, which can be insecure. To check whether the states indeed have the
form of Eq. (\ref{Bell}), they gather the outcomes when Alice did not choose
$A_{0}$, and compute the CHSH polynomial%
\begin{equation}
S=\left\langle a_{1}b_{1}\right\rangle +\left\langle a_{1}b_{2}\right\rangle
+\left\langle a_{2}b_{1}\right\rangle -\left\langle a_{2}b_{2}\right\rangle ,
\label{CHSH}
\end{equation}%
where the correlator $\left\langle a_{i}b_{j}\right\rangle $\ is defined as
the probability $P(a=b|ij)-P(a\neq b|ij)$. To generate a raw secret key,
they use the outcomes when Alice chose $A_{0}$ and Bob chose $B_{1}$. When
no eavesdropping present, there should always be $a_{0}=b_{1}$, while the
CHSH value reaches the point of maximal quantum violation of the well-known
CHSH Bell inequality, i.e., the correlators satisfy $\left\langle
a_{1}b_{1}\right\rangle =\left\langle a_{1}b_{2}\right\rangle =\left\langle
a_{2}b_{1}\right\rangle =-\left\langle a_{2}b_{2}\right\rangle =1/\sqrt{2}$\
so that $S=2\sqrt{2}$.

\section{The DI QKD protocol}

In brief, the typical structure of the DI version of the above QKD protocol
is as follows. Alice and Bob share many pairs of devices called nonlocal
boxes, which can be supplied by either Alice or Bob, or even an untrusty
third party, including the eavesdropper Eve. These devices can be treated as
DI black boxes that take inputs and produce outputs, without the need nor
the possibility to check how they work internally. Each of Alice's (Bob's)
boxes has three (two) inputs $A_{0}$, $A_{1}$\ and $A_{2}$ ($B_{1}$\ and $%
B_{2}$). For each input, Alice's (Bob's) box can product either of the two outputs $%
a_{i}=\pm 1$ ($b_{j}=\pm 1$). When no cheating present, i.e., the boxes are
manufactured honestly and working properly, Alice's and Bob's inputs to each
pair of their boxes should display the maximal nonlocality, as can be
checked by the CHSH value in Eq. (\ref{CHSH}).

In the protocol, Alice and Bob choose their inputs into the boxes randomly,
and then announce the inputs through a classical channel. Whenever Alice
chooses $A_{0}$ and Bob chooses $B_{1}$ for the boxes in the same pair, the
outputs should satisfy $a_{0}=b_{1}$ so that they can use these pairs to
generate a raw secret key. For other boxes, they gather the outcomes and
calculate the CHSH value to detect the existence of eavesdropping. See \cite%
{qi1065,qi996,qi997} for other details of the protocols.

\section{Statistical fluctuations}

It is worth noting that the maximal CHSH value $S=2\sqrt{2}$\ is a
statistical result, calculated from the mean values $\left\langle
a_{i}b_{j}\right\rangle $\ ($i,j=1,2$) of finite outcomes. Since any
statistical property is inevitably subjected to fluctuation even in the
ideal case, we cannot expect to find the actual result of an experiment
exactly equal to the theoretical value.

As a simple example, let us consider the outcomes of tossing a classical
coin. Suppose that the coin is absolutely ideal, that the outcomes
\textquotedblleft heads\textquotedblright\ and \textquotedblleft
tails\textquotedblright\ will both occur with the probability $1/2$ without
any bias at all. Even so, tossing the coin twice does not necessarily result
in $1$ head and $1$ tail. In fact this result will occur with the
probability $50\%$ only. In the rest $50\%$ case, the ratio of heads
to tails will be either $2:0$ or $0:2$. When tossing the coin $n$ times, we
cannot expect that the ratio will be perfectly $n/2:n/2$ either. In fact,
interestingly, this perfect result will occur with the probability $\binom{%
n/2}{n}/2^{n}$\ only, which drops as $n$ increases. That is, the more
samples involved in the statistics, the less we can expect to meet the
theoretical value exactly. More importantly, as we mentioned, this is the result for
an ideal coin. It remains valid even if we do not consider any experimental
imperfection, e.g., biased coins, wind disturbance, etc. Therefore, in
practice we have to accept a certain range of fluctuation deviating from the
perfect result that we expect theoretically. For instance, when tossing the
ideal coin for $10000$ times, it is still acceptable if the ratio between
heads and tails turns out to be $5500:4500$, as this result can
indeed occur with a probability not much less than that of the perfect
result $5000:5000$.

For the same reason, when checking the CHSH value in the DI QKD protocol, we
can hardly expect to find $S=2\sqrt{2}$\ even if all parties are completely
honest. Even the actual value turns out to be a little lower, the secret key
should still be considered secure. Otherwise the protocol will have only
little probability to proceed, even when no eavesdropping present at all.
Indeed, some papers suggested a precise acceptable lower bound for the CHSH
value. For example, $S\geq 2.5$ is considered acceptable in step 5 of the
protocol in \cite{qi992}.

Note that the origin of this problem is not the channel noise studied
in \cite{qi1065,di3}, whose effect is, e.g.,
following the notation in \cite{qi1065}, to transform the state Eq. (\ref%
{Bell}) into the Werner state with the density matrix%
\begin{equation}
\rho =p\left\vert \Phi ^{+}\right\rangle \left\langle \Phi ^{+}\right\vert
+(1-p)\frac{I}{4},
\end{equation}%
so that the expected CHSH value $S$ will drop. Instead, the fluctuation
considered here is a fundamental theoretical property of statistical quantities,
which exists even when the probability $p$ in the above equation vanishes.

\section{Security problems}

As $S<2\sqrt{2}$\ has to be accepted in practice, there is more room for
eavesdropping in such a DI QKD protocol, as elaborated below.

Consider that the eavesdropper Eve replaces the DI nonlocal boxes with a local
box pair $X_{1} $, whose outcomes have a fixed relationship as%
\begin{equation}
X_{1}:a_{0}=a_{1}=a_{2}=b_{1}=b_{2},  \label{X1}
\end{equation}%
where $a_{0}$ is chosen beforehand by Eve to be either $+1$ or $-1$. That
is, e.g., if $a_{0}=+1$, then Alice will obtain the outcome $a_{1}=+1$\ when
her input to her box is $A_{1}$, or she will obtain the outcome $a_{2}=+1$\
when her input is $A_{2}$. Meanwhile, Bob will obtain the outcome $b_{1}=+1$%
\ when his input to his box is $B_{1}$, etc.

Such a local box pair can be constructed, because in a DI protocol, Alice
and Bob cannot assume that the inputs $A_{i}$'s\ and $B_{j}$'s must
correspond to the operators in Eqs. (\ref{Alice}) and (\ref{Bob}). Instead,
Eve can, e.g., prepares Alice's and Bob's box pair in the product state $%
\left\vert 0\right\rangle _{A}\left\vert 0\right\rangle _{B}$ or $\left\vert
1\right\rangle _{A}\left\vert 1\right\rangle _{B}$, and let all inputs be
the operator $\sigma _{z}$. Then we can see that Eq. (\ref{X1}) can be met.

Similarly, there can be other local box pairs $X_{2}$, $X_{3}$ and $X_{4}$,
whose outcomes satisfy%
\begin{eqnarray}
X_{2} &:&a_{0}=a_{1}=-a_{2}=b_{1}=b_{2},  \nonumber \\
X_{3} &:&a_{0}=a_{1}=a_{2}=b_{1}=-b_{2},  \nonumber \\
X_{4} &:&a_{0}=-a_{1}=a_{2}=b_{1}=-b_{2},
\end{eqnarray}%
where $a_{0}$ is also chosen beforehand by Eve. These box pairs can be
constructed using non-entangled product states too. For example, $X_{2}$ can
be obtained by preparing Alice's and Bob's box pair in the three-particle product state $%
\left\vert 01\right\rangle _{A}\left\vert 0\right\rangle _{B}$, and the
inputs are set as $A_{0}=A_{1}=\sigma _{z}\otimes I$, $A_{2}=I\otimes \sigma
_{z}$, and $B_{1}=B_{2}=\sigma _{z}$.

From Eq. (\ref{CHSH}) we can easily find that the CHSH value for each of $%
X_{1}$, $X_{2}$, $X_{3}$ and $X_{4}$\ is $S=2$ as nonlocality is absent
between Alice's and Bob's boxes. Also, if Eve uses equal numbers of each of
the four box pairs, then the correlators $\left\langle
a_{1}b_{1}\right\rangle $, $\left\langle a_{1}b_{2}\right\rangle $, $%
\left\langle a_{2}b_{1}\right\rangle $, and $-\left\langle
a_{2}b_{2}\right\rangle $ in Eq. (\ref{CHSH}) all equal to $2$. Therefore no
bias can be found even if these correlators are checked separately. Now
since all the four box pairs satisfy $a_{0}=b_{1}$, whenever Alice and Bob
chooses such a pair to generate the raw secret key, Eve will know the
corresponding secret bit, while Alice's and Bob's keys remain consistent
with each other so that they find nothing wrong. The remaining question is
whether Eve can pass the CHSH value check.

Obviously, as $X_{1}$, $X_{2}$, $X_{3}$ and $X_{4}$ all have the CHSH value $%
S=2<2\sqrt{2}$, Eve cannot pass the check if she replaced all nonlocal boxes
with these four. However, as we mentioned, in the DI QKD protocol a lower
CHSH value has to be allowed due to the existence of statistical
fluctuations. Therefore if Eve uses the local boxes for a small portion, then
she will have a non-trivial probability to pass the check, while manage to
learn a part of the secret key. It turns out that the amount of the key she
learned is not too small, as we will show below.

Let $S_{\min }$ denote the lower bound of the CHSH value allowed in the DI QKD
protocol, and $S_{n}=2\sqrt{2}$ be the CHSH value expected theoretically
for the boxes displaying maximal nonlocality. Suppose that Eve
replaces the nonlocal boxes using the above local ones with the probability $%
p$. As the local boxes have the CHSH value $S=2$, while the rest nonlocal
ones that Eve has not replaced are expected to give $S_{n}=2\sqrt{2}$
averagely, the final expected value $S_{e}$ that Alice and Bob will find in
their CHSH value check is%
\begin{equation}
S_{e}=2p+S_{n}(1-p).
\end{equation}%
When $S_{e}\geq S_{\min }$\ Eve can pass the check successfully. In this case%
\begin{equation}
p\leq \frac{S_{n}-S_{\min }}{S_{n}-2}.  \label{p}
\end{equation}%
This sets the maximal probability that Eve can use the local boxes to cheat.
Since Eve will know the secret bit once a local box pair is
chosen to generate the raw key, $p$ also describes the proportion of the raw
key leaked to Eve. Substituting $S_{\min }=2.5$ (as suggested in \cite{qi992}%
) and $S_{n}=2\sqrt{2}$ into Eq. (\ref{p}), we have $p\leq 39.64\%$. As Eve
should use each of the four local boxes $X_{1}$, $X_{2}$, $X_{3}$ and $X_{4}$
with equal probabilities, this result means that she can use each box with
the probability $9.9\%$, and learn $39.6\%$\ of the raw secret key, which is
far from being trivial.

Note that for the rest $1-p=60.4\%$ nonlocal boxes that Eve has not
replaced, fluctuation also exists when calculating the CHSH value. Therefore
Eve cannot guarantee that the actual CHSH value $S_{a}$ obtained in her
above cheating will always satisfy $S_{a}>S_{\min }$. But on one hand, the fluctuation
can either raise or lower the actual value. Even if there is only about $1/2$
probability that Eve manages to learn $39.6\%$\ of the raw secret key
without being detected, it is still a serious problem to the DI QKD
protocol. On the other hand, when the local boxes $X_{1}$, $X_{2}$, $X_{3}$ and $%
X_{4} $ take part in the calculation of the CHSH value, no fluctuation will
take place as the outcomes of these boxes are deterministic. The average
fluctuation range caused by the rest $60.4\%$
nonlocal boxes is surely smaller than that in the honest protocol where $%
100\%$ of the boxes are nonlocal. If Eve uses the local boxes with a slightly
lower probability $p$, she can further raise the expected CHSH value $S_{e}$ so
that there is even less chances that the fluctuation caused by the rest
nonlocal boxes is sufficiently large to bring the actual value $S_{a}$ down below $S_{\min }$.

\section{Comparing with the DD protocol}

Intriguingly, Eve's above cheating strategy is completely futile in the
DD protocol. This is because in the DD scenario, Alice and Bob always know
what are the exact measurements that correspond to their inputs to the
states, as shown in section II. Eve has no chance to change the
measurements, even if she can replace the states. When applying the above
cheating strategy, the states she uses must be able to give a deterministic
result when Alice and Bob choose them to generate the secret key. Therefore,
the states have to be the eigenstates of Alice's measurement $A_{0}$ and
Bob's $B_{1}$. As $A_{0}=B_{1}=\sigma _{z}$\ are fixed in the DD protocol,
the states must be either $\left\vert 0\right\rangle _{A}\left\vert
0\right\rangle _{B}$\ or $\left\vert 1\right\rangle _{A}\left\vert
1\right\rangle _{B}$.

Now we calculate the CHSH values of $\left\vert 0\right\rangle _{A}\left\vert
0\right\rangle _{B}$\ and $\left\vert 1\right\rangle _{A}\left\vert
1\right\rangle _{B}$ using Eq. (\ref{CHSH}). But we can no longer set all
inputs as the operator $\sigma _{z}$ as we did in the DI case. Instead, the
operators have to remain the forms in Eqs. (\ref{Alice}) and (\ref{Bob}).
Consequently, we will find $S=\sqrt{2}<2$. But more importantly, the
correlators in $S$ will be $\left\langle a_{1}b_{1}\right\rangle
=\left\langle a_{2}b_{1}\right\rangle =1/\sqrt{2}$ while $\left\langle
a_{1}b_{2}\right\rangle =\left\langle a_{2}b_{2}\right\rangle =0$. On the
contrary, when the state $\left\vert \Phi ^{+}\right\rangle $ was used
honestly, there should be $\left\langle a_{1}b_{1}\right\rangle
=\left\langle a_{1}b_{2}\right\rangle =\left\langle a_{2}b_{1}\right\rangle
=-\left\langle a_{2}b_{2}\right\rangle =1/\sqrt{2}$. Therefore, even if Eve
mixes the states $\left\vert 0\right\rangle _{A}\left\vert 0\right\rangle
_{B}$\ and $\left\vert 1\right\rangle _{A}\left\vert 1\right\rangle _{B}$
with $\left\vert \Phi ^{+}\right\rangle $, then as long as $\left\vert
0\right\rangle _{A}\left\vert 0\right\rangle _{B}$\ and $\left\vert
1\right\rangle _{A}\left\vert 1\right\rangle _{B}$ present with a
non-trivial probability $p$, Alice and Bob will be able to find a bias on
the values of the correlators. That is, Eve cannot escape the detection if
she want to learn a non-trivial portion of the raw secret key.

\section{Summary and remarks}

Thus we can see that in the DI QKD protocol based on the violation of CHSH
inequality, a lower CHSH value has to be allowed due to the existence of
statistical fluctuation. Then the eavesdropper can have a non-trivial
probability to learn a remarkable amount of information on the raw secret
key without being detected. Note once again that this fluctuation exists
even in the ideal case. It is a fundamental theoretical problem of statistical
properties, which is not caused by any experimental imperfection. Even if
entangled states can be perfectly prepared and we can get rid of the
disturbance from the environment noise, the fluctuation will still remain.
Therefore, unfortunately, this security problem cannot be avoided by
improving the experimental technology.

Of course, by increasing the total number $n$ of the DI boxes used in the
protocol, the relative deviation from the theoretical expected value caused by the fluctuation will decrease, so that Alice and Bob can choose a higher $%
S_{\min }$ value. Therefore it will lower the ratio $p$ between the number
of Eve's obtained bits and the entire raw secret key (even though the absolute
number of bits that Eve obtained will still rise). Thus the DI QKD
protocol is still secure in the limit $n\rightarrow \infty $. However, as showed in the
previous section, the DD protocol is completely immune to the same cheating
strategy of Eve, without requiring an infinite $n$. So we
can see that quantitatively speaking, for any given finite $n$, the current
DI QKD protocol is not as secure as its DD counterpart.

The work was supported in part by the NSF of China under grant No. 10975198,
the NSF of Guangdong province, and the Foundation of Zhongshan University
Advanced Research Center.

\end{document}